\documentclass[usenatbib]{mn2e}
\pdfoutput=1
\usepackage{amssymb}
\usepackage{amsfonts}
\usepackage{amsmath}
\usepackage[pdftex]{graphicx}
\usepackage{epsfig}
\usepackage{color}
\def\twid{\mathrel{\lower.1ex\hbox{$\sim$}}}
\def\gtwid{\mathrel{\raise.3ex\hbox{$>$\kern-.75em\lower1ex\hbox{$\sim$}}}}
\def\ltwid{\mathrel{\raise.3ex\hbox{$<$\kern-.75em\lower1ex\hbox{$\sim$}}}}
\def\\{\hfil\break}

\def\eg{{\it e.g.\ }}

\def\kmSecMpc{km s$^{-1}$ Mpc$^{-1}$}

\newcommand{\be}{\begin{equation}}
\newcommand{\ee}{\end{equation}}
\newcommand{\bea}{\begin{eqnarray}}
\newcommand{\eea}{\end{eqnarray}}

\begin{document}

\title[Matter power spectrum using artificial neural networks]{PkANN -- I. Non-linear matter power spectrum interpolation through artificial neural networks}

\author[Agarwal, Abdalla, Feldman, Lahav \& Thomas]{\Large
Shankar Agarwal$^{1, \dagger}$, Filipe B. Abdalla$^{2, \S}$, Hume A. Feldman$^{1, \ddagger}$, Ofer Lahav$^{2, \flat}$ \& Shaun A. Thomas$^{2, \star}$\\
$^1$Department of Physics \& Astronomy, University of Kansas, Lawrence, KS 66045, USA.\\
$^2$Department of Physics \& Astronomy, University College London, Gower Street, London, WC1E 6BT, UK.\\
emails: $^{\dagger}$sagarwal@ku.edu, $^{\S}$fba@star.ucl.ac.uk, $^{\ddagger}$feldman@ku.edu, $^{\flat}$lahav@star.ucl.ac.uk, $^{\star}$sat@star.ucl.ac.uk
}

\date{} 

\maketitle

\begin{abstract}
We investigate the interpolation of power spectra of matter fluctuations using Artificial Neural Network (PkANN).
We present a new approach to confront small-scale non-linearities in the power spectrum of matter fluctuations. This ever-present and pernicious uncertainty is often the Achilles' heel in cosmological studies and must be reduced if we are to see the advent of precision cosmology in the late-time Universe. We show that an optimally trained artificial neural network (ANN), when presented with a set of cosmological parameters ($\Omega_{\rm m} h^2, \Omega_{\rm b} h^2, n_s, w_0, \sigma_8, \sum m_\nu$ and redshift $z$), can provide a worst-case error $\leq1$ per cent (for $z\leq2$) fit to the non-linear matter power spectrum deduced through $N$-body simulations, for modes up to $k\leq0.7\,h\textrm{Mpc}^{-1}$. Our power spectrum interpolator is accurate over the $\it{entire}$ parameter space. This is a significant improvement over some of the current matter power spectrum calculators. In this paper, we detail how an accurate interpolation of the matter power spectrum is achievable with only a sparsely sampled grid of cosmological parameters. Unlike large-scale $N$-body simulations which are computationally expensive and/or infeasible, a well-trained ANN can be an extremely quick and reliable tool in interpreting cosmological observations and parameter estimation. This paper is the first in a series. In this method paper, we generate the non-linear matter power spectra using {\sc halofit} and use them as mock observations to train the ANN. This work sets the foundation for Paper II, where a suite of $N$-body simulations will be used to compute the non-linear matter power spectra at sub-per cent accuracy, in the quasi-non-linear regime $(0.1\,h \textrm{Mpc}^{-1} \leq k \leq 0.9\,h \textrm{Mpc}^{-1})$. A trained ANN based on this $N$-body suite will be released for the scientific community.
\end{abstract}


\section{Introduction}

Studying the growth of structure and the distribution of galaxies in our Universe is a potent method for understanding both fundamental physics and the cosmological model. Measurements of the large scale structure are capable of testing our theory of gravity \citep*{AmendolaKunzSapone08}, distinguishing between dark energy models and probing the absolute neutrino mass scale \citep*{Thomas10}. It is also an immense complement to the incrementally cosmic variance limited cosmic microwave background. This complementarity explains the overwhelmingly large number of galaxy surveys on the horizon that promise to refine or even alter our understanding of the cosmos, \eg DES \citep{DES05}, the Large Synoptic Survey Telescope (LSST) \citep{LSST08}, and the Baryon Oscillation
Spectroscopic Survey (BOSS) \citep{BOSS11}. These surveys promise to achieve high-precision measurements of galaxy power spectrum amplitudes and offer a possibility to improve constraints on cosmological parameters including dark energy and neutrino masses. However, with this promise comes a great technical and systematic difficulty. 

Arguably the most ubiquitous problem in both galaxy clustering and weak lensing surveys is that as structures collapse they evolve from being linear, for which one can solve analytically, to non-linear, for which one cannot. Using $N$-body simulations \citep{Heitmann10, AgaFel10} and analytical studies inspired from perturbation theory (PT) \citep*{ScoZalHui99, Saito08}, the non-linear effects have been shown to be significant compared to the precision levels of future surveys. A consequence of this is the uncertainty in calculating the theoretical power spectra over smaller scales and at low redshift. There is frequently a choice to either exclude -- and therefore {\it waste} -- the wealth of available and expensively obtained data, or to use an inaccurate procedure, which may {\it bias} and invalidate any measurement determined with anticipated precision.

At present there are several approaches to dealing with this fruitful yet frustrating regime. One is to use sophisticated $N$-body simulations commonly produced with codes such as {\sc enzo} \citep{OShea10} and {\sc gadget} \citep{Springel05}. The most popular non-linear prescription {\sc halofit} \citep{Smith03} is a semi-analytical fit and has been fantastically successful. However, with larger and ever-improving state of the art of $N$-body simulations, the non-linearities on smaller scales have been shown to be at levels higher than the ones that were used in calibrating {\sc halofit}. As such, on smaller scales ($k\geq0.1\,h \textrm{Mpc}^{-1}$) the matter power spectra predicted by {\sc halofit} do not match the high precision $N$-body results well enough. If we are to perform precision cosmology it is imperative to go far beyond the levels of precision offered by current analytical approximations. An obstacle to further progress in obtaining accurate fits to underlying spectra is the vast computational demand from detailed $N$-body simulations and a high dimensionality in the cosmological parameter space.

There have been attempts (see \citealt*{BirdVielHae12}) to calibrate {\sc halofit} using $N$-body simulations to predict suppression of matter power spectrum for cosmological models with massive neutrinos. However, semi-analytical fits like {\sc halofit} will themselves become obsolete with near-future surveys that promise to reach per cent level of precision. Moreover, implementing neutrinos as particles in numerical simulations is a topic of ongoing research, with results (see \citealt*{Brandbyge09}; \citealt*{VieHaeSpr10}) contradictory at a level (factor of $\sim 5$ or higher) that can not be justified as due to (non)-inclusion of baryonic physics.

An alternative procedure to tackle small-scale non-linearities is to use higher order PT (\eg \citealt{Saito08}; \citealt{NisShiTarYah09}; \citealt*{Saito09}) to push further into the quasi-linear domain. Using high-resolution $N$-body simulations as reference, \cite*{CarWhiPad09} have shown that although PT improves upon a linear description of the power spectrum at large scales ($k\twid0.04\,h \textrm{Mpc}^{-1}$), it expectedly fails on smaller scales ($k\gtwid0.08\,h \textrm{Mpc}^{-1}$). The range of scales where PT is reliable at the per cent level is both redshift and cosmological model dependent. For cosmologies close to WMAP best-fit parameters, \cite{TarNisSaiHir2009} have shown that at redshift $z=0$, the one-Loop standard perturbation sequence to the non-linear matter power spectrum is expected to converge with the $N$-body simulation results to within $1$ per cent - only for scales $k\!\ltwid\!0.09\,h \textrm{Mpc}^{-1}$. With the measurements from surveys expected to be at $1$ per cent level precision, these upcoming data sets create new challenges in analyses and need alternative ways to efficiently estimate cosmological parameters.

One might argue that a machine-learning approach to determine the non-linear response from varying parameter settings is a rather black-box approach that goes against the traditional approach to spectra: based on scientific understanding and physics. However, we view this direction as a pragmatic one: a new approach is urgent given the impending flood of new data from upcoming surveys, and in an age of supposed precision cosmology, we will be theory limited in this specific area. It is therefore crucial to strive towards per cent level precision in the determination of the non-linear power spectrum.

Machine-learning techniques -- in the form of Gaussian processes -- have already been used as cosmological non-linear emulators in \citet{Habib07}, \citet{Schneider08}, \citet{Heitmann09}, \citet{Lawrence10} and \cite*{Schneider11}.  Gaussian process modeling (see \citealt{MacKay97_GP,Rasmussen06_GP} for a basic introduction to Gaussian processes) is a non-linear interpolation scheme that, after optimal learning, is capable of making predictions when queried at a suitable input setting. There are several advantages and disadvantages when using neural networks and Gaussian processes to interpolate data. From a practical point of view, a neural network compresses data into a small number of weight parameters, so a large number of simulations could be fitted into a small number of files whereas a Gaussian process has to carry a large matrix which can be of the order of the number of points used for training the Gaussian process. \citet{Heitmann09} dealt with large matrices by using principal component analysis (PCA) to reduce their sizes to ones easily manipulated. Again from a practical point of view, usually Gaussian processes can do better than neural networks in the case of a small number of training points given that a neural network could be flexible enough to be misused and misfit the data. From a theoretical point of view, the two methods should fare equally especially as there are certain kernels used in Gaussian processes which are equivalent to the interpolation and fit one would have with neural networks. Overall, given the implementation, we believe that the two methods should produce equivalent results especially if the artificial neural network (ANN) procedure is trained by a larger number of simulations.

While using any machine-learning technique in lieu of $N$-body simulation output, it is critical that (i) the queried input setting not lie outside the input parameter ranges that are used during machine learning and (ii) the input parameter space must be sampled densely enough for the machine procedure to interpolate/predict accurately. Machine learning has been used in the fitting of cosmological functions (\citealt{Auld07}; \citealt{Fendt07}; \citealt*{Auld08}) and photometric redshifts \citep{Collister04}.

{\sc halofit} is accurate at the $5-10$ per cent level at best (see \citealt{Heitmann10}). A far more accurate matter power spectrum calculator is the {\sc cosmic emulator} (see \citealt{Heitmann09}; \citealt{Lawrence10}); although accurate at sub-per cent level, it makes predictions that are valid only for redshifts $z\leq1$ and does not include cosmological models with massive neutrinos. In order to extend the interpolation validity range to $z\leq2$, as well as improve the accuracy levels, we work on a new technique to fit results from cosmological $N$-body simulations using an ANN procedure with an improved Latin hypercube sampling of the cosmological parameter space. Using {\sc halofit} spectra as mock $N$-body spectra, we show that the ANN formalism enables a remarkable fit with a manageable number of simulations.

The outline of this paper is as follows. We discuss the concept of machine learning, in particular the details of neural networks, in Section~\ref{sec:machinelearning}. The improved Latin hypercube sampling of the underlying parameter space, which keeps the simulation number manageable and fitting accuracy high, is detailed in Section~\ref{sec:lhs}.
Our fitting results are included in Section~\ref{sec:results}. Finally, we conclude in Section~\ref{sec:conclusions}.

\section{Machine Learning - The Neural Network} \label{sec:machinelearning}

Machine learning is associated with a series of algorithms that allow a computational unit to evolve in its behavior, given access to empirical data. The major benefit is the potential to automatically learn complex patterns. As a subset of artificial intelligence, machine learning has been used in a variety of applications ranging from brain-machine interfaces to the analyses of stock market. There exist a range of techniques (see \eg \citealt{Nilsson05}) including genetic algorithms, decision tree learning and gaussian processes (as referenced above). In this work we focus on the neural network technique. 

An ANN is simply an interconnection of neurons or {\it nodes} analogous to the neural structure of the brain. This can take a more specific form whereby the nodes are arranged in a series of layers with each node in a layer connected, with a weighting, to all other nodes in adjacent layers. This is often referred to as a multi-layer perceptron (MLP). In this case one can impart values onto the nodes of the first layer (called the {\it input} layer), have a series of {\it hidden} layers and finally receive information from the last layer (called the {\it output} layer). The configuration of nodes is often called the network's architecture and is specified from input to output as $N_{i}$ : $N_{1}$ : ... : $N_{n}$ : $N_{o}$. That is, a network with an architecture 7 :  49 :  50 has seven inputs, a single hidden layer with 49 nodes and finally 50 outputs. An extra node (called the {\it bias} node) is added to the input layer and connects to all the nodes in the hidden layers and the output layer (see \citealt{Bishop95} for more details). The total number of weights $N_W$ for a generic architecture $N_{i}$ : $N_{1}$ : ... : $N_{n}$ : $N_{o}$ can be calculated using the formula
\be
 \label{eq:Nodes}
N_W=N_i\cdot N_1 + \sum_{l=2}^n N_{l-1}\cdot N_l + N_n\cdot N_o + \sum_{l=1}^n N_l + N_o,
\ee
where the summation index $l$ is over the hidden layers only. For a network with a single hidden layer, the second term on the right-hand side is absent. Specifically, the architecture 7 :  49 :  50 has a total of $7\times49 + 0 + 49\times50 + 49 + 50=2892$ weights, collectively denoted by {\bf w}.

Each node (except the input nodes) is a neuron with an $\it{activation}$, $z_j=g(a)$, taking as its argument
\be
a=\displaystyle\sum_i w_{ji}z_i,
\ee
where the sum is over all nodes $i$ (of the previous layer) sending connections to the $j$th node (of the current layer). The activation functions are typically taken to be sigmoid functions such as $g(a) = 1/[1 +\rm exp(-a)]$. The sigmoid functions impart some degree of non-linearity to the neural network models. A network becomes overly non-linear if the weights {\bf w} deviate significantly from zero. This drives the activation of the nodes to saturation. The number and size of the hidden layers add to the complexity of ANNs. For a particular input vector, the output vector of the network is determined by progressing sequentially through the network layers, from inputs to outputs, calculating the activation of each node.

Adjusting the weights {\bf w} to get the desired mapping is called the {\it training} of the network. For matter power spectrum estimation, we use a training set of $N$-body simulations for which we have full information about the non-linear matter power spectra $P_{\mathrm{nl}}(k,z)$, as well as the underlying cosmological parameters: ${\bf I}\equiv(\Omega_{\rm m} h^2, \Omega_{\rm b} h^2, n_s, w_0, \sigma_8, \sum m_\nu)$, where $h, \Omega_{\rm m}, \Omega_{\rm b}, n_s, w_0, \sigma_8$ and $\sum m_\nu$ are the present-day normalized Hubble parameter in units of 100 \kmSecMpc, the present-day matter and baryonic normalized energy densities, the primordial spectral index, the constant equation of state parameter for dark energy, the amplitude of fluctuation on an 8$\,h^{-1}$ Mpc scale and the total neutrino mass, respectively. 

Given the training set, the network can be used to learn some parametrization to arbitrary accuracy by training the weights {\bf w}. This is done by minimizing a suitable $\it{cost\, function}$,
\bea
\label{eq:chisq1}
\chi^2 &=& \sum_{t=1}^T\sum_{k=1}^c \frac{[P_{\mathrm{nl}}^{\rm ANN}(k,z,{\bf w,I}_t)-P_{\mathrm{nl}}(k,z,{\bf I}_t)]^2} {\sigma(k,z,{\bf I}_t)^2}
\eea
with respect to the weights $\bf w$. The sum $t$ is over all the cosmologies ${\bf I}_t$ in the training set. The sum $k$ is over all the nodes in the output layer. Note that each output node samples the matter power spectrum at some specific scale, $k\, (h \textrm{Mpc}^{-1})$. $P_{\mathrm{nl}}(k,z,{\bf I})$ is the true non-linear matter power spectrum for the specific cosmology ${\bf I}$. In this paper, we use {\sc halofit}'s approximation for $P_{\mathrm{nl}}(k,z)$. In Paper II $N$-body simulations will be used to calculate $P_{\mathrm{nl}}(k,z)$. Given the weights {\bf w}, $P_{\mathrm{nl}}^{\rm ANN}(k,z,{\bf w,I})$ is the ANN's predicted power spectrum for the ${\bf I}$th cosmology. In our fitting procedure, we work with the ratio of the non-linear to linear power spectrum, namely $R(k,z)\equiv P_{\mathrm{nl}}(k,z)/P_{\mathrm{lin}}(k,z)$, where $P_{\mathrm{lin}}(k,z)$ is calculated using {\sc camb} \citep*{Lewis00}. As such, weighting the numerator in Eq.~\ref{eq:chisq1} by $\sigma=P_{\mathrm{lin}}(k,z)$ gives,
\bea
\label{eq:chisq2}
\chi^2 &=& \sum_{t=1}^T\sum_{k=1}^c \left[\frac{P_{\mathrm{nl}}^{\rm ANN}(k,z,{\bf w,I}_t)-P_{\mathrm{nl}}(k,z,{\bf I}_t)} {P_{\mathrm{lin}}(k,z,{\bf I}_t)}\right]^2 \\
&=&\sum_{t=1}^T\sum_{k=1}^c \left[R_{\rm ANN}(k,z,{\bf w,I}_t)-R(k,z,{\bf I}_t)\right]^2.
\label{eq:chisq3}
\eea
The ratio $R(k,z)$ is a flatter function and gives better performance, particularly at higher redshifts where the ratio tends to 1. Given the weights {\bf w}, $R_{\rm ANN}(k,z,{\bf w,I})$ in Eq.~\ref{eq:chisq3} is the network's prediction of the ratio $R(k,z,{\bf I})$ for the specific cosmology ${\bf I}$. The predicted non-linear spectrum $P_{\mathrm{nl}}^{\rm ANN}(k,z,{\bf w,I})$ in Eq.~\ref{eq:chisq2} is recovered by multiplying $R_{\rm ANN}(k,z,{\bf w,I})$ by the corresponding linear spectrum $P_{\mathrm{lin}}(k,z,{\bf I})$.

We ran $N$-body simulations over a range of cosmological parameters with the publicly available adaptive mesh refinement (AMR), grid-based hybrid (hydro+$N$-body) code {\sc enzo}\footnote{http://lca.ucsd.edu/projects/enzo} (\citealt{NorBryHarBorReySheWag07, OShea10}). We include radiative cooling of baryons using an analytical approximation \citep{Sarazin87} for a fully ionized gas with a metallicity of $0.5$ $M_{\sun}$. The cooling approximation is valid over the temperature range from $10^4-10^9$ K. Below $10^4$ K, the cooling rate is effectively zero. However, we do not account for metal-line cooling, supernova (SN) feedback or active galactic nucleus (AGN) feedback. It is worth mentioning here that \cite{VanJoop11} have shown that the inclusion of AGN feedback can reproduce the optical and X-ray observations of groups of galaxies, and decrease the power relative to dark matter-only simulations at $z= 0$, ranging from $1$ per cent at $k\approx0.4\,h \textrm{Mpc}^{-1}$ to as much as $10$ per cent at $k\approx1\,h \textrm{Mpc}^{-1}$. As such, understanding and including the effects of baryonic physics in numerical simulations will be critical to predicting the non-linear matter power spectrum at sub-per cent level. Further, the ANN prescription we are using in this paper could also be used for fitting these kinds of baryonic effects by introducing additional parameters beyond the cosmological ones, especially since gasdynamical runs are much more expensive than dark matter-only simulations.

\begin{figure}
     \includegraphics[scale=0.43]{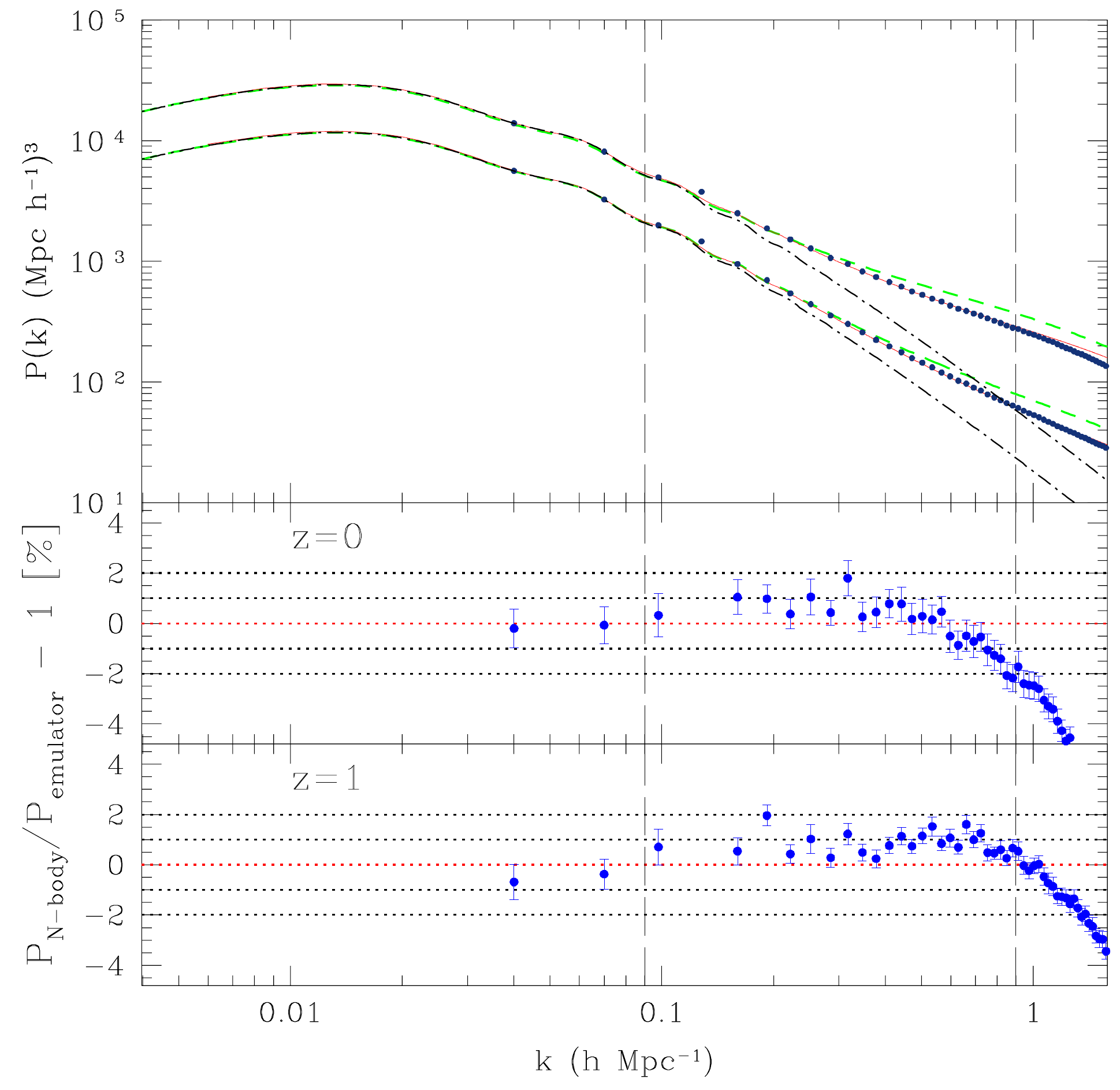}
        \caption{\small{{\it Top:} matter power spectrum evaluated at redshift $z=0$ (upper set) and $z=1$ (lower set). At each redshift, the spectrum is calculated using (i) linear theory (dot-dashed), (ii) $N$-body (dots), (iii) {\sc halofit} (dashed) and (iv) {\sc cosmic emulator} (solid, see \citealt{Lawrence10}). The cosmological parameters are: ${\bf I}\equiv(0.13, 0.0224, 0.986, -1.23, 0.72, 0)$ with $h=0.8$. In this method paper, we use the linear matter power spectrum for scales $k\leq0.09\,h \textrm{Mpc}^{-1}$. Due to the lack of force resolution on small scales, our $N$-body power spectrum is accurate at per cent level for $k\leq0.9\,h \textrm{Mpc}^{-1}$. {\it Middle:} The ratio of the $N$-body spectrum to {\sc cosmic emulator}'s prediction at $z=0$. The error bars correspond to the scatter in the $N$-body results. The horizontal dotted lines denote $\pm 2, \pm 1$ and $0$ per cent error. {\it Bottom:} The same as the middle panel, at $z=1$.
        }}
    \label{fig:enzo_coyote}
\end{figure}

\begin{figure}
     \includegraphics[scale=0.43]{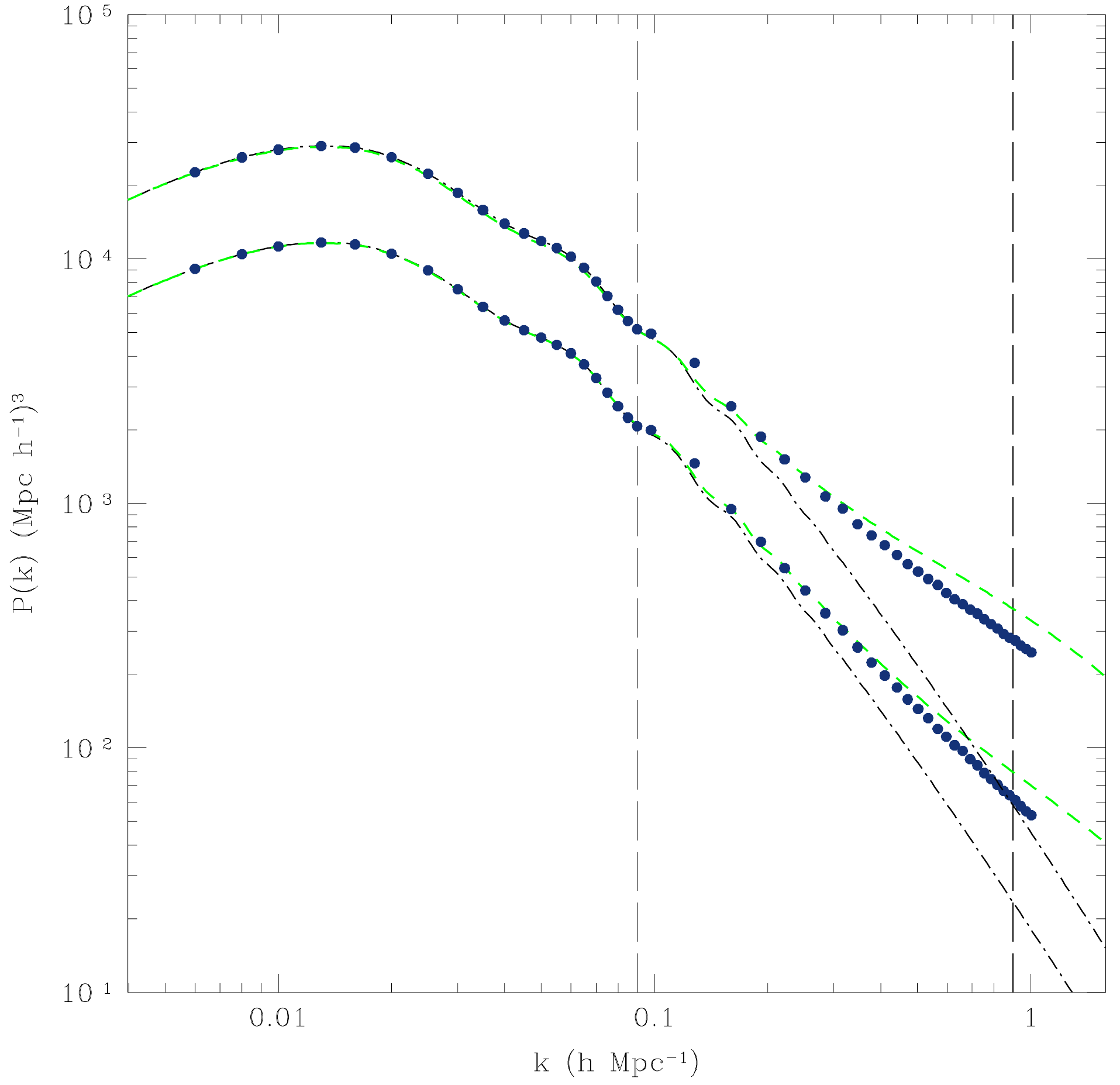}
        \caption{\small{Linear theory, {\sc halofit} and $N$-body spectra from Fig.~\ref{fig:enzo_coyote}, left-hand panel are re-plotted -- with the only difference that on scales $k\leq0.09\,h \textrm{Mpc}^{-1}$, the {\sc halofit} and $N$-body spectra are approximated by linear theory. The stitched spectra are then sampled at 50 $k$-values between $0.006\,h \textrm{Mpc}^{-1} \leq k \leq 1\,h \textrm{Mpc}^{-1}$. In this paper, we use the {\sc halofit} spectrum as $P_{\mathrm{nl}}(k,z)$ for ANN training. In Paper, II the $N$-body spectrum will be used along with the one-Loop standard PT.
        }}
    \label{fig:enzo_halofit}
\end{figure}

In Fig.~\ref{fig:enzo_coyote}, top panel, we show the power spectrum for a cosmological model ${\bf I}\equiv$ (0.13, 0.0224, 0.986, -1.23, 0.72, 0 eV), with $h=0.8$. The spectrum is evaluated at redshift $z=0$ (upper set) and $z=1$ (lower set). At each redshift, the power spectrum is calculated using (i) linear theory (dot-dashed), (ii) $N$-body (dots), (iii) {\sc halofit} (dash) and (iv) {\sc cosmic emulator} (solid). The vertical dash line at $k=0.9\,h \textrm{Mpc}^{-1}$ is the highest $k$ upto which our $N$-body power spectrum is accurate at per cent level. We average over $10$ realizations of the initial power spectrum to suppress the scatter in the $N$-body results. On smaller scales, our numerical simulations do not have the force resolution to give accurate results. The ratio of the $N$-body spectrum to {\sc cosmic emulator}'s prediction is shown in the middle ($z=0$) and bottom ($z=1$) panels. The error bars correspond to the scatter in the $N$-body results.

In this method paper, we match the linear theory power spectrum to the non-linear power spectrum from simulations at $k=0.09\,h \textrm{Mpc}^{-1}$. In Paper II we intend to use the one-Loop standard PT as implemented by \cite{Saito08} for estimating the matter power spectrum upto $k\leq0.085\,h \textrm{Mpc}^{-1}$ and stitch it with the $N$-body spectrum. The stitched spectrum will be sampled at 50 $k$-values between $0.006\,h \textrm{Mpc}^{-1} \leq k \leq 1\,h \textrm{Mpc}^{-1}$. In Fig.~\ref{fig:enzo_halofit}, we show the stitched-and-sampled $N$-body power spectrum (solid dots) which we will use as $P_{\mathrm{nl}}(k,z)$ for ANN training in Paper II. The {\sc halofit} spectrum, sampled at the same $k$-values, is also shown (open circles) and is used as $P_{\mathrm{nl}}(k,z)$ in this method paper.

In Eq.~\ref{eq:chisq3}, optimizing the weights {\bf w} in order to minimize $\chi^2$ generates an ANN that predicts the power spectrum very well for the specific cosmologies in the training set. However, such a network might not make accurate predictions for cosmologies $\it{not}$ included in the training set. This usually indicates (i) an overly simple network architecture (very few hidden layer nodes), (ii) very sparsely/poorly sampled parameter space and/or (iii) a highly complex non-linear mapping that actually over-fits to the noise on the training dataset. In order to generate smoother network mappings that generalize better when presented with new cosmologies that are not part of the training set, a penalty term $\chi^2_{Q}$ is added to the cost function $\chi^2$,
\be
\label{eq:penalty}
\chi^2_{Q} = \alpha \sum_{i,j} w_{ij}^{2},
\ee
where $w_{ij}$ is the weight connecting the $j$th node to the $i$th node of the next layer. $\chi^2_{Q}$, usually a quadratic sum of the weights, prevents them from becoming too large during the training process, by penalizing in proportion to the sum. The hyperparameter $\alpha$ controls the degree of regularization of the network's non-linearity. After having initialized $\alpha$, its value itself is re-estimated during the training process iteratively. See \cite{Bishop95} for more details.

Thus, the overall cost function which is presented to the ANN for minimization with respect to the weights $\bf w$ is,
\be
\label{eq:cost}
\chi^2_{C}=\sum_{t=1}^T\sum_{k=1}^c \left[R_{\rm ANN}(k,z,{\bf w,I}_t)-R(k,z,{\bf I}_t)\right]^2 + \alpha \sum_{i,j} w_{ij}^{2}.
\ee

To minimize the cost function $\chi^2_{C}$ w.r.t. the weights {\bf w}, we use an iterative quasi-Newton algorithm which involves evaluating the inverse Hessian matrix
\be
\label{eq:Hess}
H_{ij}=\frac{\partial^2  \chi^2_{C}}{\partial w_i w_j}.
\ee

Specifically, we employ the Broyden--Fletcher--Goldfarb--Shanno (BFGS) method to approximate the inverse Hessian matrix (see \citealt{Bishop95}, for details). Starting with randomly assigned weights {\bf w}, their values are re-estimated iteratively, assuring that each iteration proceeds in a direction that lowers the cost function $\chi^2_{C}$. After each iteration to the weights, Eq.~\ref{eq:cost} is also calculated for what is known in neural network parlance as a validation set, in order to avoid over-fitting to the training set. The validation set for our application of neural networks, is a small set of simulations with known ${\bf I}\equiv(\Omega_{\rm m} h^2, \Omega_{\rm b} h^2, n_s, w_0, \sigma_8, \sum m_\nu)$ and $P_{\mathrm{nl}}(k,z)$. The final weights ${\bf w}_f$ are chosen such as to give the best fit (minimum $\chi^2_{C}$) to the validation set. The network training is considered finished once $\chi^2_{C}$ is minimized with respect to the validation set. The trained network can now be used to predict $P_{\mathrm{nl}}(k,z)$ for new cosmologies. In practice, a number of networks are trained that start with an alternative random configuration of weights. The trained networks are collectively called a committee of networks and subsequently give rise to better performance. The final output is usually given by averaging over the outputs of the committee members.

The ANN technique has been used successfully in empirical photometric redshift estimation with the {\sc annz} \citep{Collister04} package. {\sc annz} learns an effective parameterization of redshift as a function of broad-band photometric colours by training on a representative set of galaxies that have both photometric and spectroscopic information. This has been shown to be more successful than template or synthetic-based methods (\citealt{Abdalla11}; \citealt*{Thomas11}). In this work we use an MLP similar to the original {\sc annz} engine.

Our intention is to use this neural network technique to learn the non-linear matter power spectrum as a function of cosmological parameters by training on $N$-body simulations. This natural fitting procedure removes both the effort and unnecessary potential bias that results from invoking ultimately imperfect sets of fitting equations such as the {\sc halofit}. As can be seen in Fig.~\ref{fig:enzo_halofit}, the {\sc halofit} predictions are in error by as much as $50$ per cent on small scales. As we will discuss in Section~\ref{sec:results}, the ANN technique is extremely fast and, more importantly, accurate.

\section{Latin Hypercube Parameter Sampling} \label{sec:lhs}

In order to fit a set of parameters optimally one strives to sample them as finely and as evenly as possible. However, a regularly spaced grid with $N$ sampling intervals along one dimension and $d$ parameters scales as $N^{d}$. For a six-dimensional parameter space with only $10$ grid intervals, this quickly escalates to $10^6$ points. The problem is exacerbated because an $N$-body simulation is a computationally expensive activity. To further compound this issue, each parameter configuration needs to be simulated over multiple realizations to beat down simulation (sample) variance. An alternate approach could be to interpolate the fitting function over a selection of randomly distributed points throughout the parameter space. However, this is prone to statistical clustering and will lead to a degradation of the machine-learning fit for parameters or regions affected by it. In order to circumvent these problems, we select parameters distributed on a Latin hypercube.
 
\begin{figure*}
  \begin{flushleft}
   \centering
    \begin{minipage}[c]{1.00\textwidth}
      \centering
      \includegraphics[width=5.5cm,height=5.5cm]{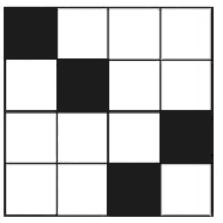}
      \includegraphics[width=5.5cm,height=5.5cm]{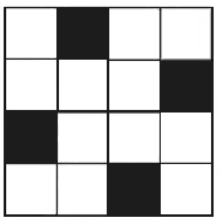}
    \end{minipage}
    \caption{\small{{\it Left:} An example of a Latin hypercube distribution. Every interval $\mathrm{d}x$ and $\mathrm{d}y$ is sampled; however each row and column are sampled only once. {\it Right:} an improved Latin hypercube where the distribution is more evenly spread through the space. Each subspace is equally sampled and there are no voids or clusters as in the left-hand panel (bottom left and right corners, respectively).
    }}
    \label{fig:latin}
  \end{flushleft}
\end{figure*}

A square grid is said to be populated as a Latin square if, and only if, there is exactly {\it one} sample in each row and each column of the square. This is illustrated clearly in Fig.~\ref{fig:latin}. A similar sampling scheme was developed first by Leonhard Euler who indexed the samples with Latin characters, motivating the name `Latin square'. A Latin hypercube is a generalization of Euler's Latin square to a higher dimensional parameter space and is an example of a stratified sampling technique. This ensures that each and every segment/interval along a parameter axis is sampled with high resolution without a vast number of points. That is, one can sample a $d$-dimensional space with $n$ simulations and have all parameters evaluated along every $\mathrm{d}x = (b-a)/n$ increment, where $b$ and $a$ are the upper and lower limits of the parameter, respectively. Therefore, it is independent of $d$. However, a random implementation of a Latin hypercube algorithm can still lead to statistically under-sampled regions.  An example of this can be seen in Fig.~\ref{fig:latin}. Each panel shows a random implementation of Latin hypercube algorithm. In both panels, the square is partitioned into four subspaces. The left-hand panel has voids (and clusters) in two of its subspaces. The right-hand panel has each subspace equally sampled (while still obeying the Latin hypercube definition) and represents an improved Latin hypercube sampling. In this case the sample space is partitioned into equally probable subspaces and the variance in the pairwise separation of the sampled points is minimized.

Since the introduction of the Latin hypercube sampling technique \citep{McKay79}, the procedure has become common in computer science, uncertainty analysis and engineering emulation (where simulation of complex machinery is overwhelmingly time consuming). Similarly, variations of the Latin hypercube sampling technique have been implemented in cosmological analyses before, e.g., \citet{Habib07}, \citet{Heitmann09}, \citet{Schneider11} and references therein. In this paper, we use the improved Latin hypercube technique to set up the cosmological models to be used for ANN training.

\subsection{Setting up an improved Latin hypercube for cosmological parameters}

We varied six cosmological parameters ${\bf I}\equiv(\Omega_{\rm m} h^2, \Omega_{\rm b} h^2, n_s, w_0, \sigma_8, \sum m_\nu)$ between the limits specified in Table~\ref{tab:train_set_priors}. The limits on this six-dimensional parameter space are chosen so as to include the WMAP 7-year+BAO+$H_0$ \citep{Komatsu11} constraints (see Table~\ref{tab:train_set_priors}).
\begin{table*}
\caption{\small{The six cosmological parameters and their ranges, used in generating the ANN training and validation sets. This six-dimensional parameter space is sampled using the improved Latin hypercube technique (see text for details). The last column shows the corresponding WMAP 7-year+BAO+$H_0$ constraints at 68 per cent CL.}}
\begin{tabular}{cccc} \hline
 \multicolumn{1}{c}{Cosmological parameters}& \multicolumn{1}{c}{Lower value} &  \multicolumn{1}{c}{Upper value} &  \multicolumn{1}{c}{WMAP 7-year+BAO+$H_0^a$} \\  \hline
 $\Omega_{\rm m}h^2$		& 0.110 	& 0.165	& 0.1352		$\pm$ 0.0036	\\
 $\Omega_{\rm b}h^2$		& 0.021 	& 0.024	& 0.02255		$\pm$ 0.00054	\\
 $n_{\rm s}$				& 0.85 	& 1.05	& 0.968		$\pm$ 0.012	\\
 $w_0$					& -1.35	& -0.65	& -1.1		$\pm$ 0.14	\\
 $\sigma_8$				& 0.60	& 0.95	& 0.816		$\pm$ 0.024	\\
 $\sum m_\nu$ (eV)			& 0		& 1.1		& $<0.58^b$			\\ \hline
 \multicolumn{1}{l}{{\it Note.} $^a$\cite{Komatsu11}; $^b95\,$ per cent CL for $w=-1.$}
\end{tabular}
\label{tab:train_set_priors}
\end{table*}

Throughout this paper, we only consider spatially flat models with the present-day CMB temperature ${T_\gamma}^0=2.725K$. We also assume that all massive neutrino species are degenerate. The effective number of neutrino species is fixed at $N_{\rm eff}=3.04$. We derive the Hubble parameter $h$ using the WMAP 7-year+BAO constraint on the acoustic scale ${\pi d_{ls}}/{r_s}=302.54$, where $d_{ls}$ is the distance to the last scattering surface and $r_s$ is the sound horizon at the redshift of last scattering. We derive $h$ as follows.

(i) For a given $\Omega_b h^2$ and $\Omega_m h^2$, compute the redshift of the last scattering surface, $z_{ls}$, using the fit proposed by \cite{HuSu96}:
\bea
\label{eq:z_ls}
z_{ls}&=&1048\left[1+\frac{0.00124}{(\Omega_b h^2)^{0.738}}\right]\left[1+b_1 (\Omega_m h^2)^{b_2}\right]\\
b_1&=&\frac{0.0783}{(\Omega_b h^2)^{0.238}}\left[1+39.5(\Omega_b h^2)^{0.763}\right]^{-1}\\
b_2&=&\frac{0.560}{1+21.1(\Omega_b h^2)^{1.81}}
\eea

(ii) For a given $\Omega_b h^2$, $\Omega_m h^2$ and $\sum m_\nu$, choose a value for $h$ and compute its evolution, $h(a)$. Here we follow section 3.3 from \cite{Komatsu11}, which includes the effect of massive neutrinos on $h(a)$:
\bea
\label{eq:h_a}
h(a)&=&h\sqrt{\frac{\Omega_b+\Omega_c}{a^3}+\frac{\Omega_\gamma}{a^4}\left[1+\frac{\Omega_\nu}{\Omega_\gamma}\right]+\frac{\Omega_\Lambda}{a^{3(1+w_0)}}}\\
\frac{\Omega_\nu}{\Omega_\gamma}&=&N_{\rm eff}\frac{7}{8}\left(\frac{4}{11}\right)^{4/3}F(y)\\
F(y)&=&\frac{120}{7\pi^4}\int_{0}^{\infty} \frac{x^2\sqrt{x^2+y^2}}{e^x+1}dx,
\eea
where
\bea \nonumber 
y&\equiv&\frac{m_{\nu}a}{{T_\nu}^0}\\ \nonumber
{T_\nu}^0&=&\left(\frac{4}{11}\right)^{1/3}{T_\gamma}^0\\ \nonumber
\Omega_{\gamma}&=&\frac{2.4706\times10^{-5}}{h^2}\left(\frac{{T_\gamma}^0}{2.725}\right)^4.
\eea

${T_\nu}^0$ is the present-day neutrino temperature and $\Omega_\gamma$ is the present-day normalized photon energy density. Given $\sum m_\nu$, the function $F(y)$ calculates the contribution of neutrinos to the radiation energy density at scale factor $a$.

(iii) Using $h(a)$ from step (ii), compute the comoving sound horizon $r_s(z)$ at the last scattering redshift $z_{ls}$:
\bea
\label{eq:r_ls}
r_s(z_{ls})&=& \frac{c}{\sqrt{3}} \int_{a=0}^{1/(1+z_{ls})} \frac{da}{a^2 h(a) \sqrt{1+(3\Omega_b / 4\Omega_\gamma)a}}.
\eea

(iv) Using $r_s(z_{ls})$ from step (iii), together with the WMAP 7-year+BAO constraint on the acoustic scale ${\pi d_{ls}}/{r_s}=302.54$, compute the comoving distance to the last scattering surface, $d_{ls}$:
\bea
d_{ls}=\frac{302.54}{\pi}r_s(z_{ls}).
\eea

(v) Using $h(a)$ from step (ii), compute the comoving distance to the surface of last scattering $\chi(z_{ls})$:
\bea
\chi(z_{ls})&=&c\int_{1/(1+z_{ls})}^{a=1} \frac{da}{a^2h(a)}.
\eea

(vi) Compare results from steps (iv) and (v). Minimize the difference $|d_{ls}-\chi(z_{ls})|$ by varying $h$ in step (ii) and re-estimating steps (ii)-(v).\\

\begin{figure}
     \includegraphics[scale=0.47]{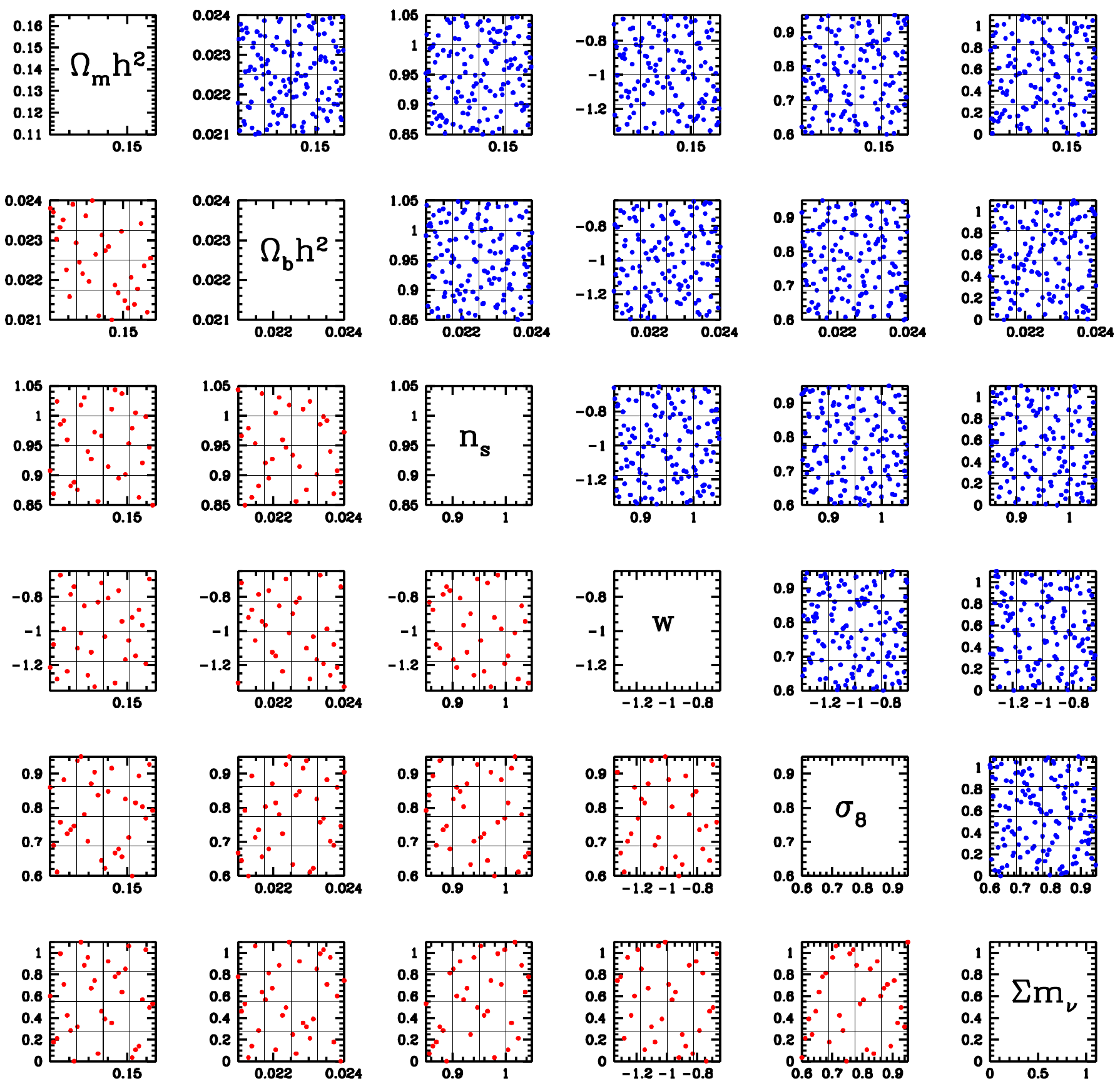}
        \caption{\small{Upper triangle: ANN training set with 130 viable cosmologies, in a six-dimensional parameter space. Lower triangle: ANN validation set with 32 viable cosmologies, in a six-dimensional parameter space. See Table~\ref{tab:train_set_priors} for the parameter priors used to generate the training and validation sets.
        }}
    \label{fig:LHS_train_valid}
\end{figure}

Using Table~\ref{tab:train_set_priors} as the parameter priors, we sampled this six-dimensional parameter space with an improved Latin hypercube technique.  We generated 130 cosmologies to be used as the ANN training set and another 32 cosmologies for the validation set. We show the training set (upper triangle) and the validation set (lower triangle) in Fig.~\ref{fig:LHS_train_valid}.

\begin{figure}
     \includegraphics[scale=0.47]{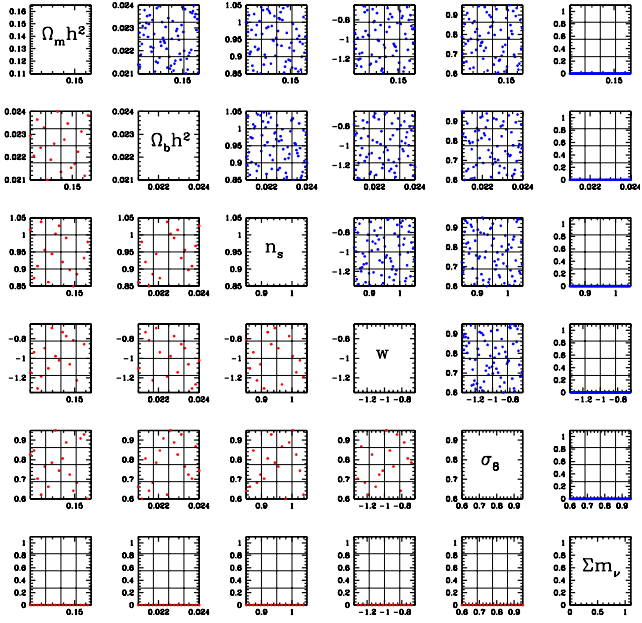}
        \caption{\small{Upper triangle: extending the ANN training set (upper triangle in Fig.~\ref{fig:LHS_train_valid}) with 70 cosmologies with $\sum m_{\nu}=0$. Lower triangle: extending the ANN validation set (lower triangle in Fig.~\ref{fig:LHS_train_valid}) with 18 cosmologies with $\sum m_{\nu}=0$.
        }}
    \label{fig:LHS_train_valid_0eV}
\end{figure}

As can be seen in Fig.~\ref{fig:latin}, a major advantage of improved Latin hypercube sampling technique is the relatively uniform coverage it provides. This is, of course, highly useful for training a machine-learning algorithm. As with any interpolation mechanism, one hopes that the neural network can generalize from what it has learned to new and slightly different input data (in this case cosmological parameters). In reality, the response will be uncertain in poorly trained areas. Therefore, the caveat with our sampling will reside near the edges of the parameter hypercube. A parameter value that we might want emulated may not be encapsulated within the hypervolume of a simulated, and therefore trained, point. This can be understood with reference to Fig.~\ref{fig:LHS_train_valid}. The performance of a neural network can severely degrade near the parameter boundaries. The solution is simply to choose prior ranges that are marginally wider than those of real interest. The allowance could easily be found empirically by projecting the hypercube realizations. The real problem in cosmology therefore arises when one has a parameter that is physically bounded, an example being the neutrino mass $\sum m_{\nu} \gtrsim 0$.

Adding several additional simulations at the parameter boundary may not be a computationally feasible solution to the problem due to the multi-dimensionality of the parameter space. Instead we propose to use a nested hypercube with $6-1=5$ dimensions. We fixed $\sum m_{\nu}=0$ and varied the rest of the parameters over their aforementioned limits. We extended the ANN training and validation sets to include this five-dimensional hyperplane. Towards this, we generated 70 (for training) and 18 (for validation) cosmologies with $\sum m_{\nu}=0$. Fig.~\ref{fig:LHS_train_valid_0eV} shows the five-dimensional hyperplane.




\section{Analysis and Results} \label{sec:results}

We tested the precision with which our neural network can predict the non-linear matter power spectrum. We selected the combination 7 :  $N_{hidden}$ :  50 as our network architecture, where $N_{hidden}$ (number of nodes in the hidden layer) was varied from 7 to 56, in steps of 7. The number of inputs were fixed at 7, corresponding to ${\bf I}\equiv(\Omega_{\rm m} h^2, \Omega_{\rm b} h^2, n_s, w_0, \sigma_8, \sum m_\nu)$ including redshift $z$. Note that we do not sample the redshift in the Latin hypercube but instead evaluate $P_{\mathrm{nl}}(k,z)$, at 111 redshifts between $z=0$ and $z=2$, using the existing prescription {\sc halofit} coupled with the {\sc camb} software. These {\sc halofit}-generated spectra serve as mock $N$-body spectra.  We sampled $P_{\mathrm{nl}}(k,z)$ at 50 points between $0.006\,h \textrm{Mpc}^{-1} \leq k \leq 1.0\,h \textrm{Mpc}^{-1}$. Since our training and validation sets have ($130+70$) and ($32+18$) cosmologies, respectively, we calculated $P_{\mathrm{nl}}(k,z)$ for each cosmology, at 111 redshifts. These $P_{\mathrm{nl}}(k,z)$ are scaled by their respective linear spectra $P_{\mathrm{lin}}(k,z)$, before being fed to the neural network. Thus, the overall size of the training set that we train our ANN with is $200\times111=22,200$. Likewise, we have $50\times111=5,550$ patterns in the validation set. We trained a committee of 16 ANNs at each $N_{hidden}$ setting. The weights $\bf w$ for each ANN were randomly initialized (the random configuration being different for each ANN). The weights are allowed to evolve until $\chi^2_{C}$ (see Eq.~\ref{eq:cost}) is minimized with respect to the cosmologies in the validation set.

\begin{figure}
     \includegraphics[width=85mm,height=100mm]{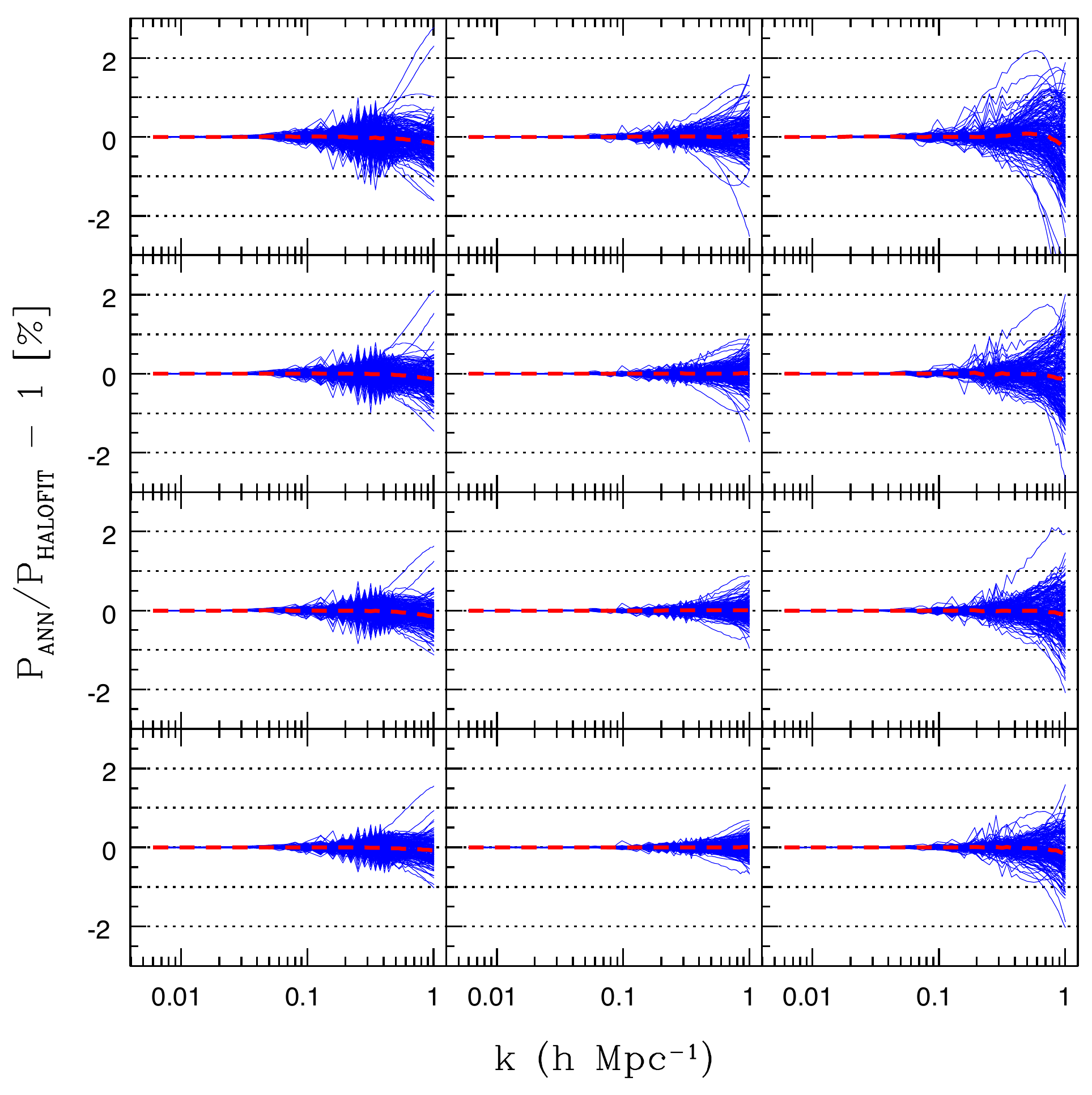}
        \caption{\small{{\it Left-hand column:} percentage error at redshift $z=0$, between the predicted non-linear power spectrum (using ANN) and the true underlying mock (using {\sc halofit}) for $200$ training set cosmologies shown in the upper triangles of Figs.~\ref{fig:LHS_train_valid} and~\ref{fig:LHS_train_valid_0eV}. The profile is continuous as the 50 output values have been spline interpolated over the functional range. The rows (from top to bottom) correspond to $N_{hidden}=28-49$ in increments of 7. The mean error over all 200 cosmologies is shown by a dashed line -- an indicator of any bias in the ANN training scheme. {\it Middle- column:} the same as the left-hand column, at redshift $z=1$. {\it Right-hand column:} the same as the left-hand column, at redshift $z=2$.
    }}
    \label{fig:fit_error_training}
\end{figure}

In Fig.~\ref{fig:fit_error_training}, we show the performance of the trained ANNs with varying $N_{hidden}$ units, when presented with each of the 200 cosmologies in the training set. Note that we average the $P_{\mathrm{nl}}^{\rm ANN}(k,z)$ predictions over all 16 ANN committee members. The rows correspond to $N_{hidden}=28-49$ (from top to bottom) in increments of 7. The columns (from left to right) correspond to $z=0,1,2$. The mean error over all 200 cosmologies in the training set is shown by a dashed line in each panel, to get an idea about any systematics in our ANN training scheme. With $N_{hidden}=49$, the ANN predictions at redshifts $z=0$ and $z=1$, on $\it{all}$ scales, are within $\pm 1$ per cent of the {\sc halofit} power spectra. Although we show results at $z=0$ and $z=1$, we have checked that the predictions are $1$ per cent level for all $z\leq1$. Predictions are at the $1$ per cent level even up to redshift $z=2$ for $k\leq0.7\,h \textrm{Mpc}^{-1}$, after which the performance degrades to $\pm 2$ per cent. We have checked and confirmed that the worst-performing cosmologies correspond to the parameter settings in which at least four of the six cosmological parameters are at their boundary values.

As we mentioned earlier, this fitting procedure will be less accurate near the boundaries of the parameter ranges because some parameter configurations may not be encapsulated within the volume of a training point. This also explains why the ANN performance is better at $z=1$ -- the mid-point of the redshift range. Looking at the bias (dashed line in Fig.~\ref{fig:fit_error_training}), we see that the distribution of errors in the ANN predictions is centered on zero, indicating that our interpolations are not biased. A negligible bias, and the fact that for $\it{every}$ cosmological setting $\it{within}$ the parameter priors (see Table~\ref{tab:train_set_priors}) the non-linear power spectrum at $z\leq2$ is correctly predicted within $\pm 1$ per cent up to $k\leq0.7\,h \textrm{Mpc}^{-1}$, demonstrates the stability of our ANN strategy. This marks a remarkable improvement over the currently popular interpolation scheme -- the {\sc cosmic emulator}, which has a significant number ($\twid\!50$ per cent) of cosmological models with errors at $\twid\!0.5-1$ per cent level. We note, however, that the {\sc cosmic emulator}, based on Gaussian processes, is able to achieve sub-per cent accuracy with only 37 distinct cosmologies while in the ANN scheme we use a suite of around 200 cosmologies. Comparing Fig. 9 from \citet{Heitmann09} with our Fig.~\ref{fig:fit_error_training}, we see that the ANN implementation performs better on all scales and redshifts.

In order to check the performance of our trained ANNs over parameter configurations that do not touch the Latin hypercube, we generated a testing set of 330 cosmologies (of which 150 have $\sum m_\nu=0$). See Table~\ref{tab:test_set_priors} for the parameter limits of the testing set.
\begin{table*}
\caption{\small{The six cosmological parameters and their ranges, used in generating the ANN testing set. This six-dimensional parameter space is sampled using the improved Latin hypercube technique (see the text for details). The parameter ranges are chosen so as to avoid the boundaries of the parameter space. See Table~\ref{tab:train_set_priors} for the parameter boundaries. Note that the lower bound on neutrino mass is still set at zero, since neutrinos are physically bound ($\sum m_{\nu} \gtrsim 0$). The last column shows the WMAP 7-year+BAO+$H_0$ constraints at 68 per cent CL.}}
\begin{tabular}{cccc} \hline
 \multicolumn{1}{c}{Cosmological parameters}& \multicolumn{1}{c}{Lower value} &  \multicolumn{1}{c}{Upper value} &  \multicolumn{1}{c}{WMAP 7-year+BAO+$H_0^a$} \\  \hline
 $\Omega_{\rm m}h^2$		& 0.120 	& 0.150	& 0.1352		$\pm$ 0.0036	\\
 $\Omega_{\rm b}h^2$		& 0.022 	& 0.023	& 0.02255		$\pm$ 0.00054	\\
 $n_{\rm s}$				& 0.90 	& 1.00	& 0.968		$\pm$ 0.012	\\
 $w_0$					& -1.15	& -0.85	& -1.1		$\pm$ 0.14	\\
 $\sigma_8$				& 0.70	& 0.85	& 0.816		$\pm$ 0.024	\\
 $\sum m_\nu$ (eV)			& 0		& 0.50	& $<0.58^b$			\\ \hline
 \multicolumn{1}{l}{{\it Note.} $^a$\cite{Komatsu11}; $^b95\,$ per cent CL for $w=-1.$}
\end{tabular}
\label{tab:test_set_priors}
\end{table*}

\begin{figure}
     \includegraphics[scale=0.47]{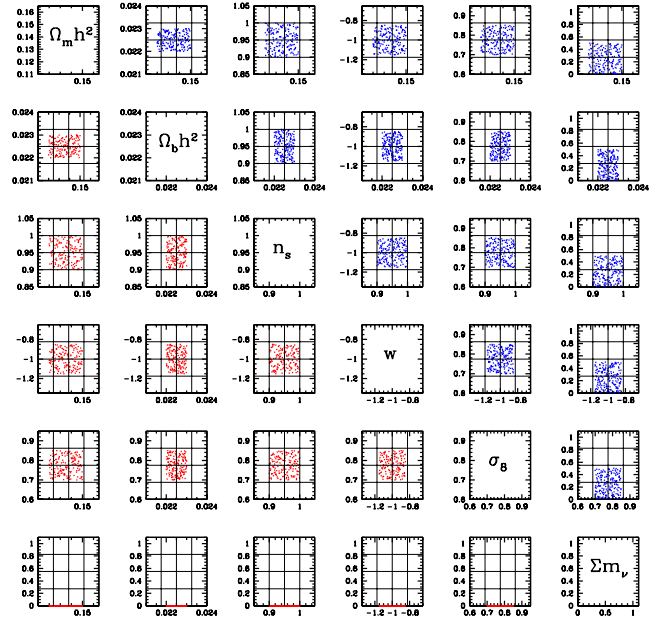}
        \caption{\small{Upper triangle: ANN testing set with 180 cosmologies with $\sum m_{\nu}>0$. Lower triangle: extending the ANN testing set with 150 cosmologies with $\sum m_{\nu}=0$. See Table~\ref{tab:test_set_priors} for the parameter priors used to generate the testing set.
        }}
    \label{fig:LHS_test}
\end{figure}

A testing set serves another crucial purpose. Increasing the number of nodes in the hidden layer increases the flexibility of a neural network. An increasingly complex network can make extremely accurate predictions on the training set. This is evident from Fig.~\ref{fig:fit_error_training}, where the prediction over the training set becomes progressively better (from top to bottom) with increasing $N_{hidden}$ units. However, such complex networks can adversely affect their generalizing ability when presented with a new dataset. Measuring the performance of a neural network on an independent dataset (known as a testing set) as a function of $N_{hidden}$ units helps in controlling its complexity. We show the testing set in Fig.~\ref{fig:LHS_test}, with the lower triangle corresponding to the 150 cosmologies with $\sum m_\nu=0$.

\begin{figure}
     \includegraphics[width=85mm,height=100mm]{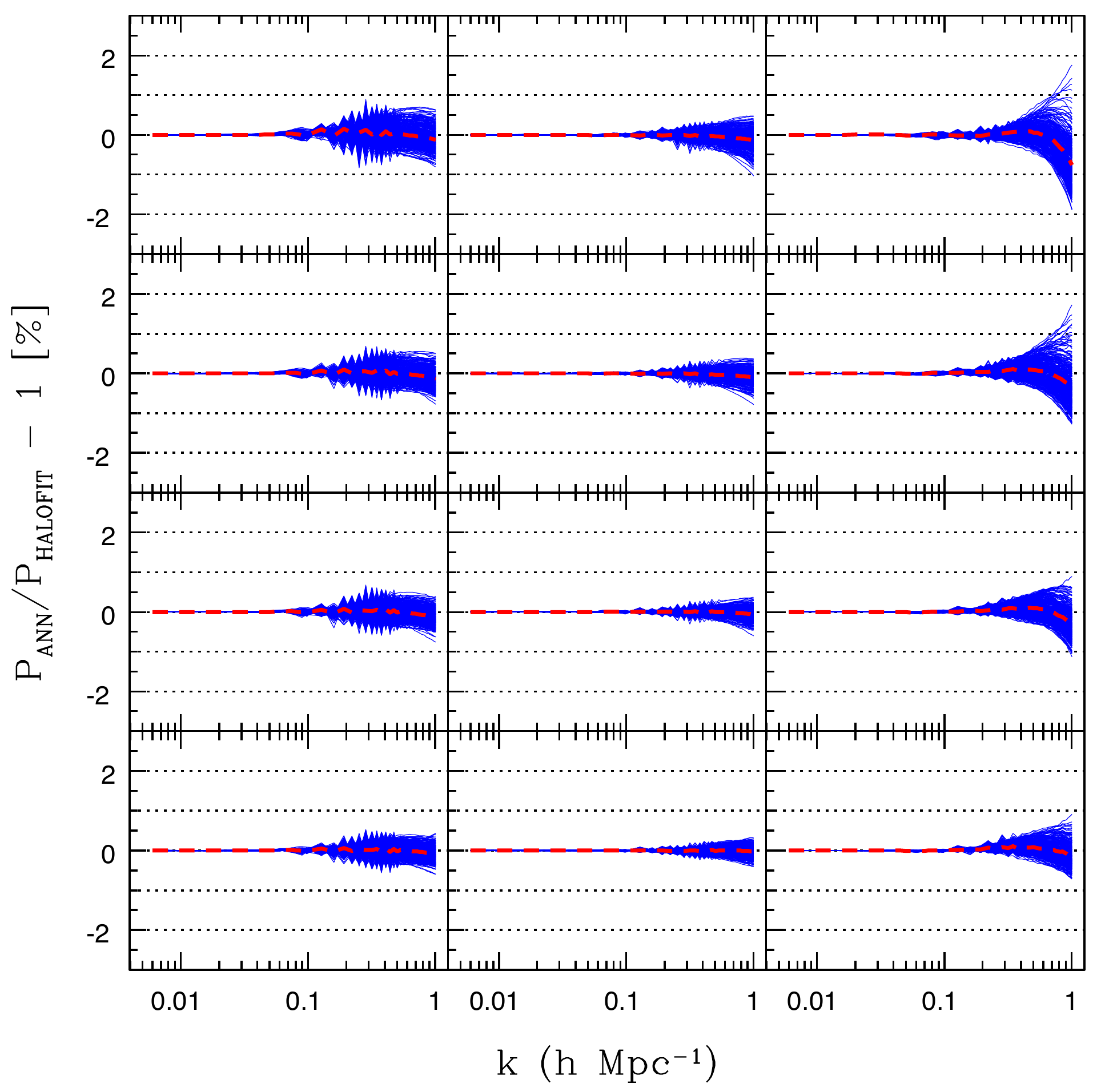}
        \caption{\small{The same as Fig.~\ref{fig:fit_error_training}, using 330 testing set cosmologies shown in Fig.~\ref{fig:LHS_test}.
    }}
    \label{fig:fit_error_testing}
\end{figure}

The performance of the trained ANNs as a function of $N_{hidden}$ units, over the cosmologies in the testing set, is shown in Fig.~\ref{fig:fit_error_testing}. Increasing $N_{hidden}$ from 42 to 49 reduces the error marginally. We have checked that increasing $N_{hidden}$ beyond 49 does not contribute to any further error reduction on the testing set, indicating that $N_{hidden}=49$ saturates the generalizing ability of the network. With $N_{hidden}=49$, the ANN prediction for every cosmology, on $\it{all}$ scales, at redshifts $z\leq1$, is within $\pm 0.5$ per cent of the {\sc halofit} power spectra. For $1<z\leq2$, in the high $k$ range, the performance degrades slightly to $\pm 0.8$ per cent. As expected, the ANN performs exceedingly well away from the boundaries of the parameter ranges. It is quite remarkable that our ANN scheme is capable of making predictions at sub-per cent level, especially on the testing set that is $\it{not}$ a part of the ANN training process.

\begin{figure*}
  \begin{flushleft}
   \centering
    \begin{minipage}[c]{1.00\textwidth}
      \centering
      \includegraphics[width=5.7cm,height=5.7cm]{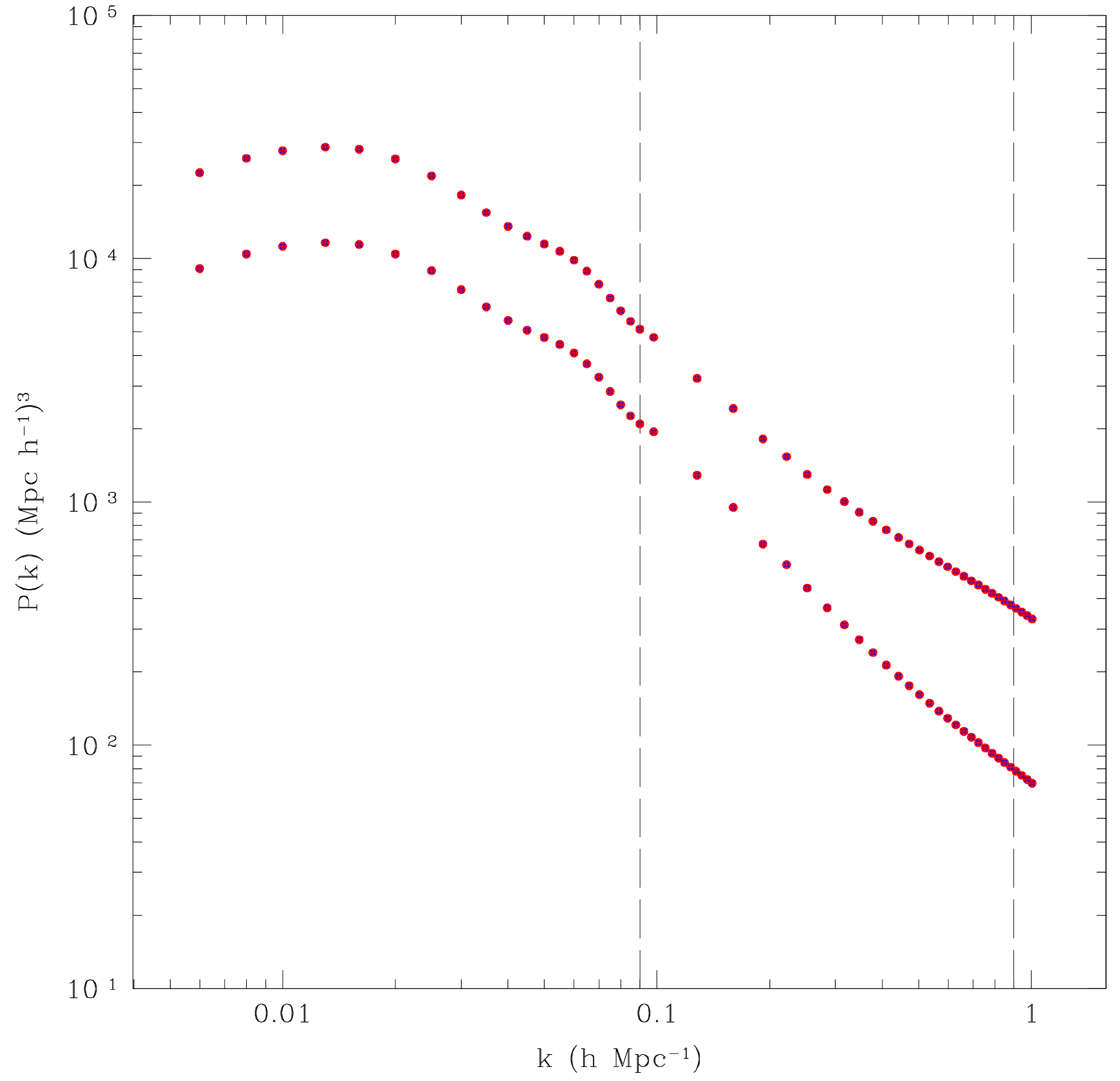}
      \includegraphics[width=5.7cm,height=5.7cm]{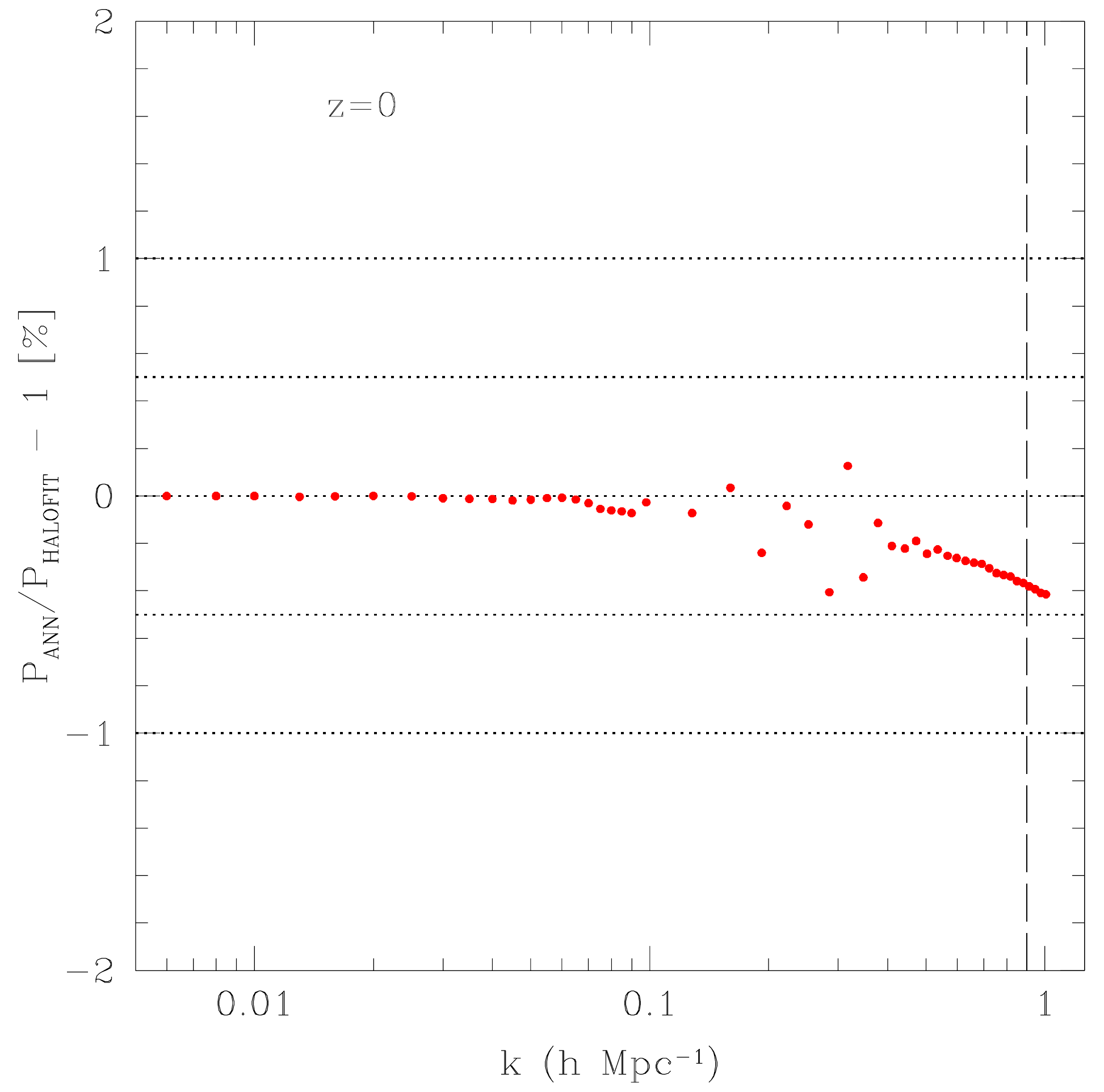}
      \includegraphics[width=5.7cm,height=5.7cm]{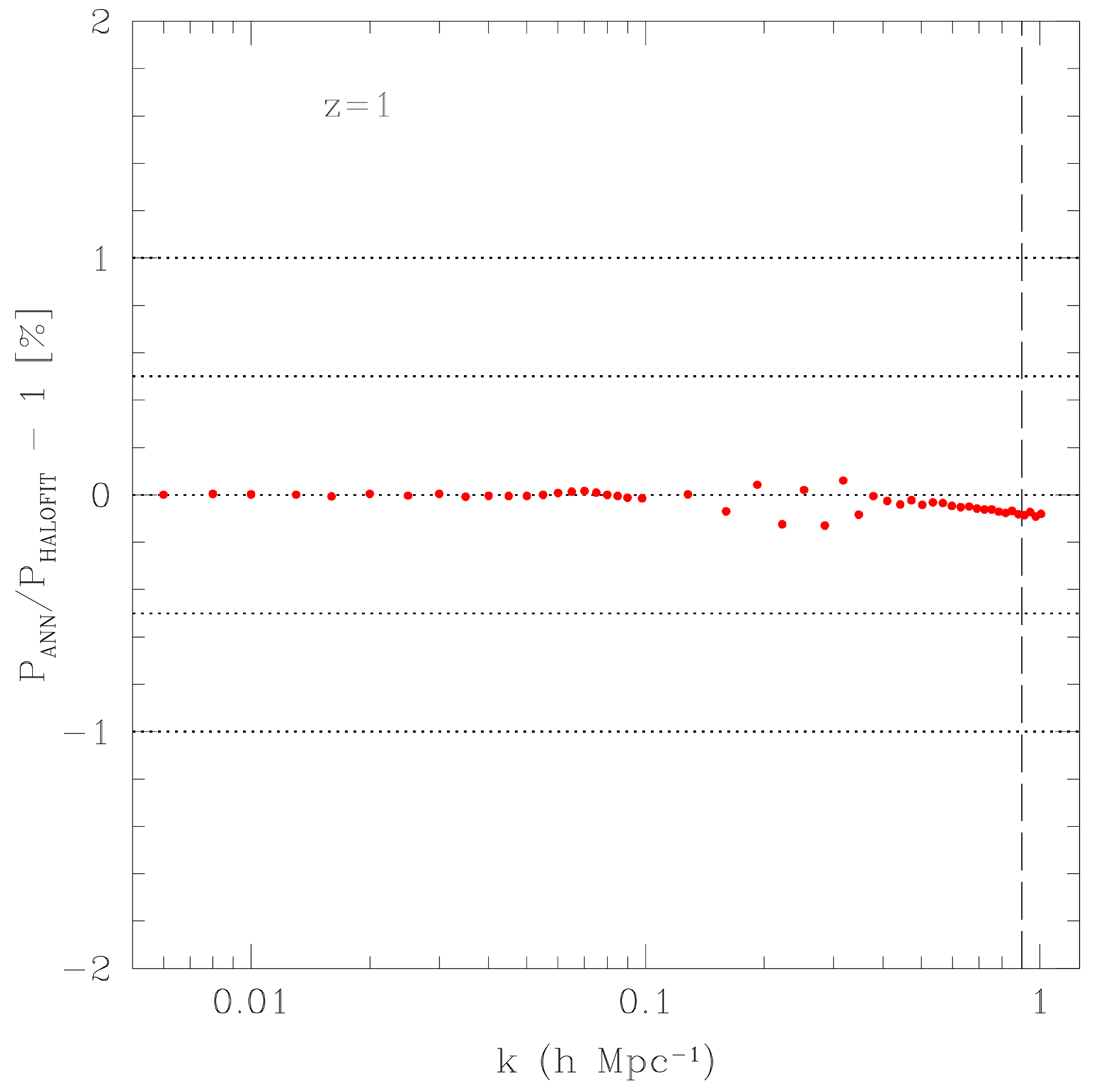}
    \end{minipage}
    \caption{\small{{\it Left:} ANN predictions (solid dots) at $z=0$ (upper) and $z=1$ (lower) for the {\sc halofit} power spectra of Fig.~\ref{fig:enzo_halofit} (re-plotted here as starred points -- difficult to see due to the excellent {\sc ann} predictions). The cosmological model is ${\bf I}\equiv$ (0.13, 0.0224, 0.986, -1.23, 0.72, 0 eV) with $h=0.8$. $N_{hidden}$ is fixed at 49, as discussed in the text. The ratio of {\sc ann} predictions to the {\sc halofit} spectra is shown in the other panels. {\it Middle:} percentage error in ANN predictions, at $z=0$. {\it Right:} percentage error in ANN predictions, at $z=1$. The ANN predictions are within $\pm 0.5$ per cent over the scales of interest.
    }}
    \label{fig:Pann_Phalofit_pk}
  \end{flushleft}
\end{figure*}

In Fig.~\ref{fig:Pann_Phalofit_pk}, left panel, we show the non-linear power spectrum at redshifts $z=0$ and $z=1$ predicted by our trained ANN (solid dots) for the same cosmology that was used to generate the {\sc halofit} power spectrum of Fig.~\ref{fig:enzo_halofit}, namely ${\bf I}\equiv$ (0.13, 0.0224, 0.986, -1.23, 0.72, 0 eV) with $h=0.8$. $N_{hidden}$ is fixed at 49, as discussed above. For comparison, the {\sc halofit} power spectra are re-plotted as starred points. The prediction errors at $z=0$ and $z=1$ are shown in the middle and the right-hand panels, respectively. The ANN predictions are well within $\pm 0.5$ per cent over the scales of interest. 

We reiterate that this method for reconstructing the non-linear power spectrum will only function for the parameters and ranges that have been simulated and trained with {\sc PkANN}. The intention of this study is to provide a technique for high precision fits in the concordance model for the oncoming generation of surveys. This should therefore act as a safety mechanism as it demonstrates that the range of validity has been breached, as often occurs with blind application of other fits. In Paper II, we will put our ANN formalism to test using matter power spectra calculated from $N$-body simulations. On mildly non-linear scales ($0.1\,h \textrm{Mpc}^{-1} < k < 1.0\,h \textrm{Mpc}^{-1}$) the power spectrum from $N$-body simulations is expected to vary smoothly as a function of cosmological parameters \citep{Heit06, Heitmann09, AgaFel10}. As such, our ANN interpolation scheme that we have shown here to work well on {\sc halofit} power spectra is expected to perform satisfactorily on $N$-body spectra as well.

\section{Conclusions} \label{sec:conclusions}

The advent of the era of precision cosmology poses an immense challenge to theoretical physics. The upcoming generation of surveys has the potential to breach per cent level of accuracy. Such high-precision data will improve our constraints on cosmological parameters including dark energy and neutrino masses. Efficiently dealing with this impending flood of precise data on ever smaller scales and lower redshifts requires that we move on from linear theory as well as any imperfect sets of fitting equations. Although numerical simulations are capable of achieving the levels of precision required by the near-future surveys, the high dimensionality of the cosmological parameter space renders their brute force usage impractical.

We have introduced a unique approach to coping with non-linearities in the matter power spectra in cosmology. By employing a multi-layer perceptron neural network together with an improved Latin hypercube parameter sampling technique, we have demonstrated that the non-linear spectrum can be reconstructed from a full set of $\Lambda$ cold dark matter parameters to better than $1$ per cent over the parameter space spanning $3\sigma$ confidence level around the WMAP 7-year central values. Parameters that are likely to reside by some hard physical prior, such as the neutrino mass, can be successfully brought under the realms of ANNs by sprinkling of extra simulations in the corresponding (\eg $\sum m_{\nu}=0$) hyper-plane.

Looking forward, our ANN procedure can be readily employed for a variety of cosmological tasks such as fitting halo mass functions obtained through high resolution $N$-body simulations. Moreover, mixed datasets such as the matter power spectra and the halo mass functions can be combined and presented to a neural network as the training set. An ANN trained with such a heterogeneous dataset would be capable of cosmological parameter estimation when presented with the combined observations of the matter power spectrum and the measured halo mass function. The implementation of our technique avoids complex calculations and, through the execution of only the neural network weights, is extremely fast (predictions take less than a second). An automated {\sc PkANN} function will be released with our program of $N$-body results in a companion paper shortly. Beyond this we hope that with our method a collaborative effort could reduce non-linear error to only uncertainty in the $N$-body simulation's baryon interactions.


\noindent
\section{Acknowledgements}

We thank Salman Habib, Katrin Heitmann and Joop Schaye for their insightful comments and suggestions on the present paper. This work is supported by the National Science Foundation through TeraGrid resources provided by the NCSA. SAT acknowledges UCL's Institute of Origins for a Post-doctoral Fellowship. FBA and OL acknowledge the support of the Royal Society via a Royal Society URF and a Royal Society Wolfson Research Merit Award, respectively.

\bibliographystyle{mn2e}
\bibliography{pkann}

\end{document}